\documentclass[twocolumn,showpacs,preprintnumbers,amsmath,amssymb]{revtex4}
\usepackage [latin1]{inputenc}
\usepackage{graphicx}
\usepackage{dcolumn}
\usepackage{bm}

\begin{document}
\title[]{Geometrical and electrical modulation on the transport property of silicene constrictions}
\author{Yawen Guo}

\author{Wenqi Jiang}
\author{Xinru Wang}
\author{Yijing Bai}
 \author{Fei Wan}
  \author{Guanqing Wang}
   \email{gqwang@hdu.edu.cn}
 \author{Yuan Li}
 \email{liyuan@hdu.edu.cn}
\affiliation{
Department of Physics, Hangzhou Dianzi University, Hangzhou, Zhejiang 310018, China}

\begin{abstract}
We study the electrical modulation of the transport properties of silicene constrictions with different geometrical structures by adopting the tight-binding model and non-equilibrium Green's function method. The band structure and transmission properties are discussed under the influence of the external electric field and potential energy. Especially, we investigate the effects of the position and width of the central scattering region on the conductance with increasing of Fermi energy. We find that the conductance significantly depends on the position and the width. Interestingly, the symmetrical structure of the central region can induce a resonance effect and significantly enlarge the system's conductance. Obviously, we obtain an effective method to adjust the transport property of the silicene heterojunctions. Correspondingly, we propose a novel two-channel structure with an excellent performance on the conductance compared to the one-channel structure with the same total width.
\end{abstract}
\date{\today}
\maketitle

\section{\label{sec:level1}INTRODUCTION}

Silicene has a low-buckle monolayer-honeycomb structure formed from a monolayer silicon atoms. In the recent years, after being synthesized on metal surfaces successfully\cite{aufray2010graphene,lalmi2010epitaxial,de2010evidence}, it has attracted much attention between researchers of both theoretical\cite{cahangirov2009two,fagan2000ab} and experimental fields\cite{chen2012evidence,vogt2012silicene}. Its low-buckled geometry creates a relatively large gap opened by the spin-orbit coupling at Dirac points\cite{Liu2011}. The most common substrate for silicene is Ag(111) surface, which is investigated fully and verified that the substrate-induced symmetry breaking will annihilate the Dirac electrons near Fermi level in silicene \cite{guo2013,Lin2013}. It is also reported that the size of band gap increases as the external electric field strengthen and a phase transition from a topological insulator to a band insulator will happen in the process\cite{Ezawa_2012}. What's more, silicene stimulates the development of many fields involved with valley-polarized quantum anomalous Hall effect\cite{Ezawa_2012,pan2014valley}, quantum spin Hall effect\cite{liu2011quantum,tabert2013ac}, spin and valley polarization\cite{missault2015spin,stille2012optical}, topologically protected edge states\cite{Ezawa2013,Cano-Cortes}, etc.
\begin{figure}[b]
\centering
\includegraphics[scale=0.58]{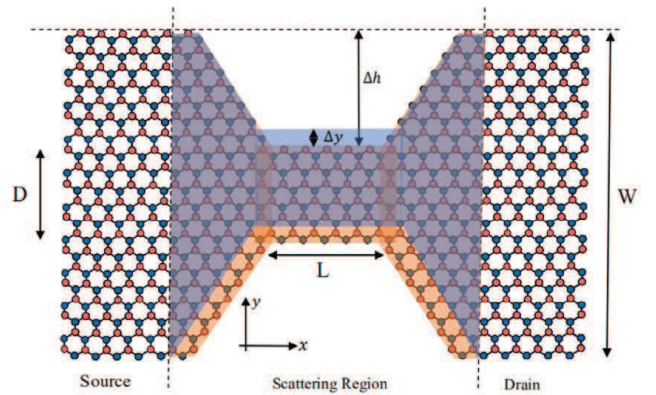}
\caption{Schematic of the silicene constriction. The zigzag direction is parallel to the $x$ axis as it's shown in the picture.
The orange region shows a central symmetrical scattering region for the silicene constriction.
The blue part gives a different geometry structure with $\Delta y$ referring to the relative distance of the top edge
to that of the orange geometry.}
\label{fig:move structure}
\end{figure}

In order to make silicene a better candidate for valleytronics devices, many researches are devoted to modulate the band gap and transmission conductance. Specifically, researchers are trying to control the energy band by applying an external electric field\cite{van2014spin}, a strain\cite{li2019electrically} or a gate voltage\cite{yamakage2013charge,guzman2018transport}. The valley and spin separation can be achieved when a strain and an electric field are simultaneously applied to silicene\cite{li2018strain}. In addition, it has been reported that the modulation of transmission  of silicene nanoribbons by changing the lead\cite{gu2018effects}. And the self-similar transport beyond graphene has been achieved through metallic electrodes arranged in Cantor-like fashion over silicene\cite{rodriguez2020self}. However, the effect of geometrical modulation in silicene devices has not been discussed extensively. In this paper, we discuss the effect of the position and width of the central scattering region on the transport property with the external field and potential energy.

\section{THEORETICAL MODEL}
Silicene is made of a hexagonal honeycomb lattice of silicon atoms, and its two sublattices is usually marked by $A$ and $B$ sites. The distance of two sublattice planes is $a_z=0.23\mathring{A}$. The silicene nanoribbon is put on a Cartesian coordinate system with the zigzag direction along $x$ axis with the lattice constant being $a=3.86\mathring{A}$. Fig.\ref{fig:move structure} shows a silicene device whose scattering region can situate at different positions.

We define the original position to be in the middle, making the entire structure symmetrical, as shown in the orange part in the figure. The relative distance of the top edge of the blue part to that of the orange geometrical structure is $\Delta y=M\xi$, in which $\xi=\frac{\sqrt{3}}{2}a=3.34~\mathring{A}$ and M is integer ($M=0,\pm1,\pm2,\pm3,...)$ representing the number of sites for the relative shift. We calculate the conductance of the silicene constrictions with both the positive and negative value of M, finding that the results are equal when the absolute values of M are equal. This is reflected in the geometric structure that the positive and negative values of M are symmetrical. Therefore, we will only discuss the case where M is positive, that is, $M=0, 1, 2, 3,...$. The length of the scattering region is 800 sites and the wide part has 40 sites. That is $L\approx153nm, W\approx13nm$. The width of the central region is $D=(N-1)\xi+\frac{1}{4}a$, in which $N=2, 3, 4,...$ is the sites number of the central narrow part.

We analyse the transport properties of this silicene constrictions under the presence or absence of the external electric field and the voltage potential, respectively. We adopt the four-band second-nearest-neighbour tight-binding model to describe this device, whose Hamiltonian can be written as the following form~\cite{Liu2011, Kane2005, Motohiko_Ezawa}:

\begin{multline}
H=-\sum_{i\alpha}a_z\mu_i E_z c_{i\alpha}^\dagger c_{i\alpha}+\sum_{i\alpha}V_i c_{i\alpha}^\dagger c_{i\alpha}
-\epsilon\sum_{\langle i,j\rangle\alpha} c_{i\alpha}^\dagger c_{j\alpha}
\\+i\frac{t_{SO}}{3\sqrt{3}}\sum_{\langle\langle i,j\rangle\rangle\alpha\beta} \kappa_{ij}c_{i\alpha}^\dagger \sigma_{\alpha\beta}^z c_{j\beta}
\\+i\frac{2}{3}t_{R}\sum_{\langle\langle i,j\rangle\rangle\alpha\beta} \mu_i c_{i\alpha}^\dagger(\boldsymbol{\sigma}\times\boldsymbol{\hat{d}}_{ij})_{\alpha\beta}^z c_{j\beta}.
\label{hamiltonian}
\end{multline}
The first term associates with the staggered sublattice potential. The low-buckled structure leads to a distance $a_z$ between two sublattices made of $A$ sites and $B$ sites, hence produces the sublattice potential, which is proportional to product of $a_z$ and the electric field $E_z$, in which $\mu_i=\pm 1$ for the $A$ (B) site. $c_{i\alpha}^\dagger$ ($c_{i\alpha}$) represents the creation (annihilation) operator with spin index $\alpha$ at site $i$. The second term is on-site potential energy induced by gate voltage. The third term describes the hopping of the nearest neighbors, and the transfer energy is $\epsilon=1.09~\mathrm{eV}$. $ \langle i,j \rangle$ and $\langle\langle i,j\rangle\rangle$ mean traversing all the nearest- or next-nearest-neighbor hopping sites.
 The fourth term is relevant to the effective spin-orbit coupling concerning the hopping between the next-nearest-neighboring sites. The coupling coefficient is $t_{SO}\thickapprox 3.9 ~\mathrm{meV}$ and $\boldsymbol{\sigma}=(\sigma_x, \sigma_y, \sigma_z)$ is the Pauli matrix.
$\kappa_{ij}$ decides the hopping path for what $\kappa_{ij}=1$ means anticlockwise respected to the positive direction of the $z$ axis, and $\kappa_{ij}=-1$ does the opposite. The last term represents Rashba spin-obit coupling between next-nearest-neighboring sites with $t_R=0.7 \mathrm{meV}$.
\begin{figure}[t]
\centering
\includegraphics[scale=0.5]{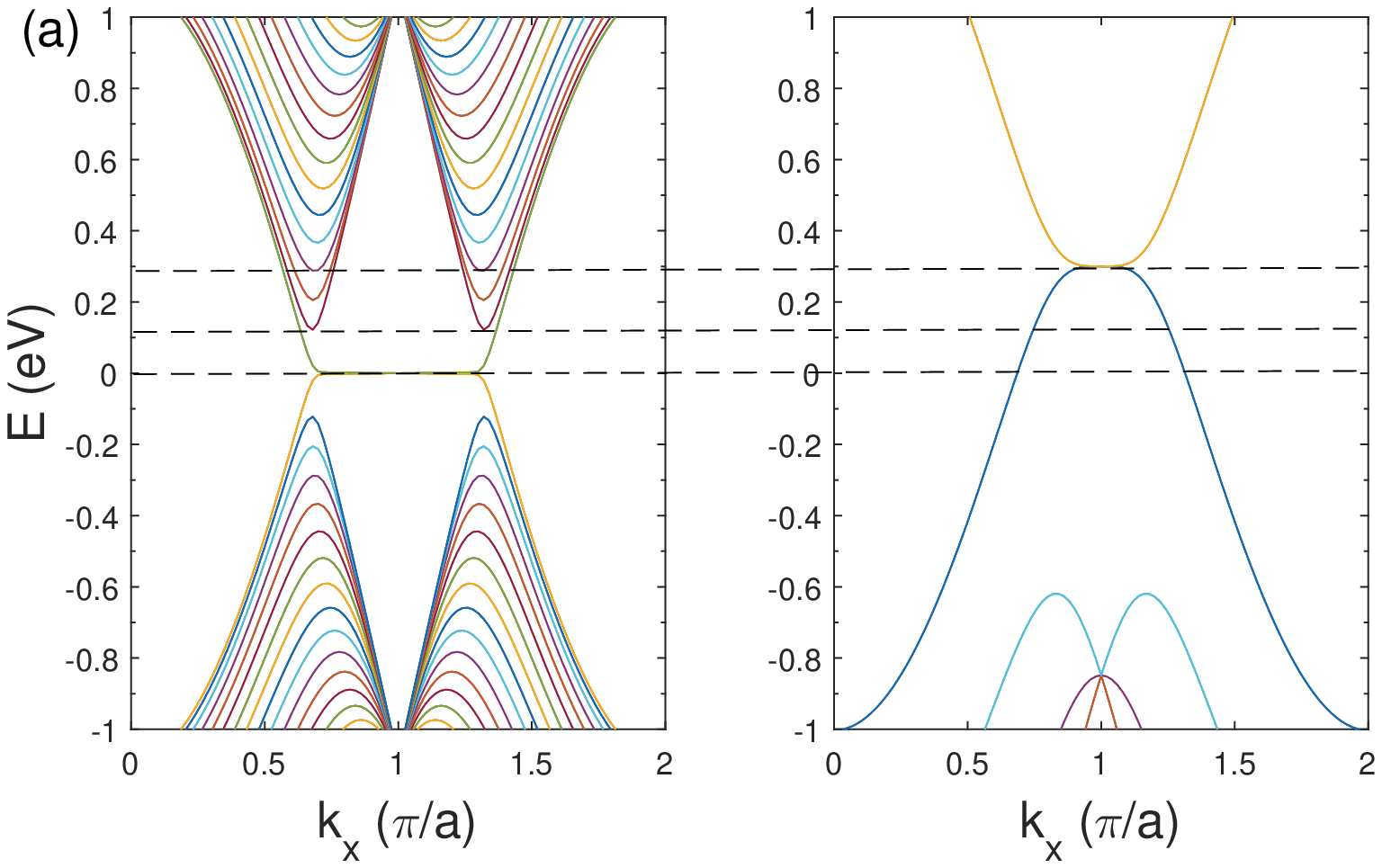}
\includegraphics[scale=0.5]{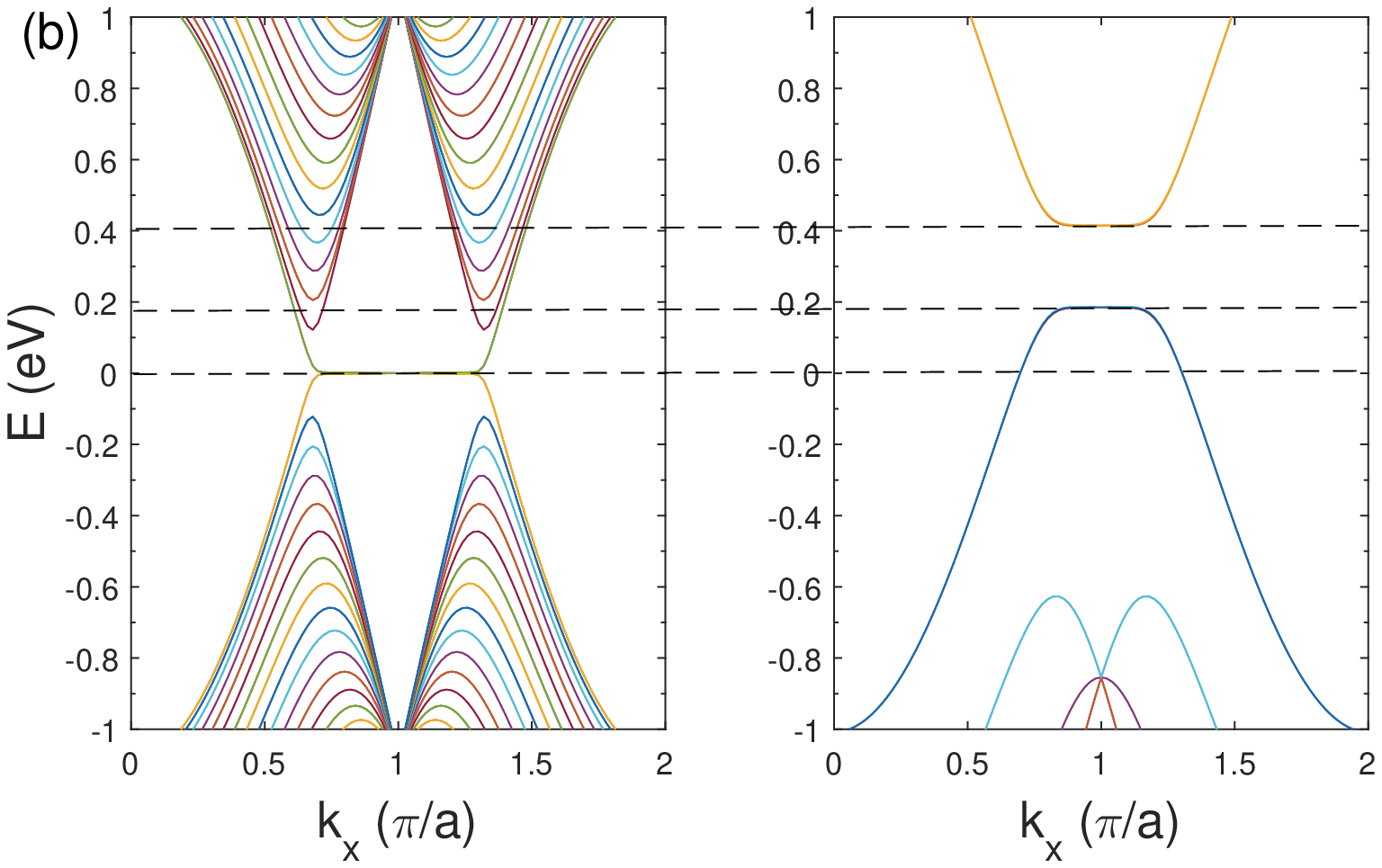}
\caption{The band structure of the lead (left) and the central scattering region (right) with the electric field (a) $E_z=0$ and (b) $E_z=0.5~\mathrm{eV{\AA}^{-1}}$. The potential energy $V_0=0.3~\mathrm{eV}$ and the sites number of the central narrow part is $N=4$.}
\label{fig:band energy}
\end{figure}

The conductance is calculated by the Landauer-B\"uttiker formalism, which is also called scattering formalism, relating the conductance to the properties of the scattering wave function. It describes the linear conductance between two leads for small bias voltages and
low temperatures, i.e. in the linear response regime. The conductance is proportional to the sum of the transmission probabilities between the corresponding modes ($n,m$) of the leads. It can be written as\cite{Buttiker1985, wimmer2009quantum}:
\begin{equation}
    G=\frac{e^2}{h}\sum_{mn}|t_{n,m}|^2,
\end{equation}
where $t_{n,m}$ represents the transmission amplitude from mode m to n. The Schr\"odinger equation of the silicene constriction is
\begin{eqnarray}
-\mathbf{H}_{\tau,\tau-1}\psi_{\tau-1}+(E\mathbf{I}-\mathbf{H}_{\tau,\tau})\psi_{\tau}-\mathbf{H}_{\tau,\tau+1}\psi_{\tau+1}=0,
\end{eqnarray}
where $\tau=-\infty,\ldots, 0$ and $\tau=S+1, \ldots, +\infty$ refer to the cells of left and right leads, $\tau=1, 2,\ldots, S$
associates with the central scattering region. $\psi_\tau$ is a N-dimensional vector for cell $\tau$, and $\mathbf{H}_{\tau,\tau'}$ denotes the $N\times N$ Hamiltonian matrix referring to the hopping terms between sites of cells $\tau$
and $\tau'$, which can be obtained from the tight-binding Hamiltonian in Eq.~(\ref{hamiltonian}).

One can first obtain the wave functions of leads by solve the generalized eigenvalue equation\cite{li2018strain, Ando1991}.
The wave functions can be divided into propagating and evanescent modes with the mode number $n=1,2, \ldots, N$.
We can obtain the transmission matrix elements $t_{n,m}$ by solving the Schr\"odinger equation of the scattering regions when we expand the renormalized vector $\psi_{S+1}(+)$ in the modes of the right lead,
\begin{eqnarray}
\psi_{S+1}(+)=\sum_n^N \psi_{R,n}(+)t_{n,m},
\end{eqnarray}
where $\pm$ refers to the right-going and left-going propagating modes, $\psi_{R/L,n}(+)$ is the n-th right-going propagating mode of the right (left) lead.
Consequently, the transmission matrix elements $t_{n,m}$ can be written as:
\begin{equation}
    t_{n,m}=\Tilde{\psi}_{R,n}^\dagger(+)G_{S+1,0}[G_{0,0}^{(0)}]^{-1}\psi_{L,m}(+).
\end{equation}
where $G_{0,0}^{(0)}$ and $G_{S+1,0}$ are the Green's functions of the left lead and the full system. Since
the effect of the leads has been treated as the boundary conditions, the indexes $\tau=0, S+1$ refer to the left lead and right lead, respectively. $\Tilde{\psi}_{n}^\dagger(\pm)$ are dual vectors, which satisfy the following relations
\begin{eqnarray}
\Tilde{\psi}_{n}^\dagger(\pm)\psi_{m}(\pm)=\delta_{n,m}, ~~~\psi_{n}^\dagger(\pm)\Tilde{\psi}_{m}(\pm)=\delta_{n,m}.
\end{eqnarray}
The Green's functions $G_{0,0}^{(0)}$ and $G_{S+1,0}$ can be obtained by using the iterative techniques of the Green's function method~\cite{Khomyakov2005}.

\section{RESULTS AND DISCUSSION}
We mainly investigated the effects of structural changes such as the position and width of the central scattering region combined with the electric field and potential energy on the transport properties. First of all, we discussed the dispersion relationship of silicene with zigzag edges in the presence of electric field $E_z$ and potential energy. Secondly, we study the influence of the positional deviation of the central nanoribbon on the conductance of the silicene heterojunction with or without an external electric field and potential energy. Thirdly, we study the effect of the width of the central region on the conductance. At last, we proposed a two-channel structure and found that we could obtain a higher conductance with this structure.

Fig.~\ref{fig:band energy}(a) shows the band structures of the lead (left) and the central scattering region (right) with the electrical field $E_z=0$ and $V_0=0.3 \mathrm{meV}$. It can be illustrated by Fig.~\ref{fig:band energy}(b) that when applying an electrical field $E_z=0.5~\mathrm{eV{\AA}^{-1}}$, a band gap will emerge. It could be generalized from the results that the band structure and gap will be changed by the potential energy and electric energy respectively. On this basis, we continue our numeral calculations to study how the geometry of the central scattering zone will influence the transport properties of silicene in the following passages.
\begin{figure}[t]
\centering
\begin{minipage}[b]{0.48\linewidth}
\includegraphics[width=1\linewidth]{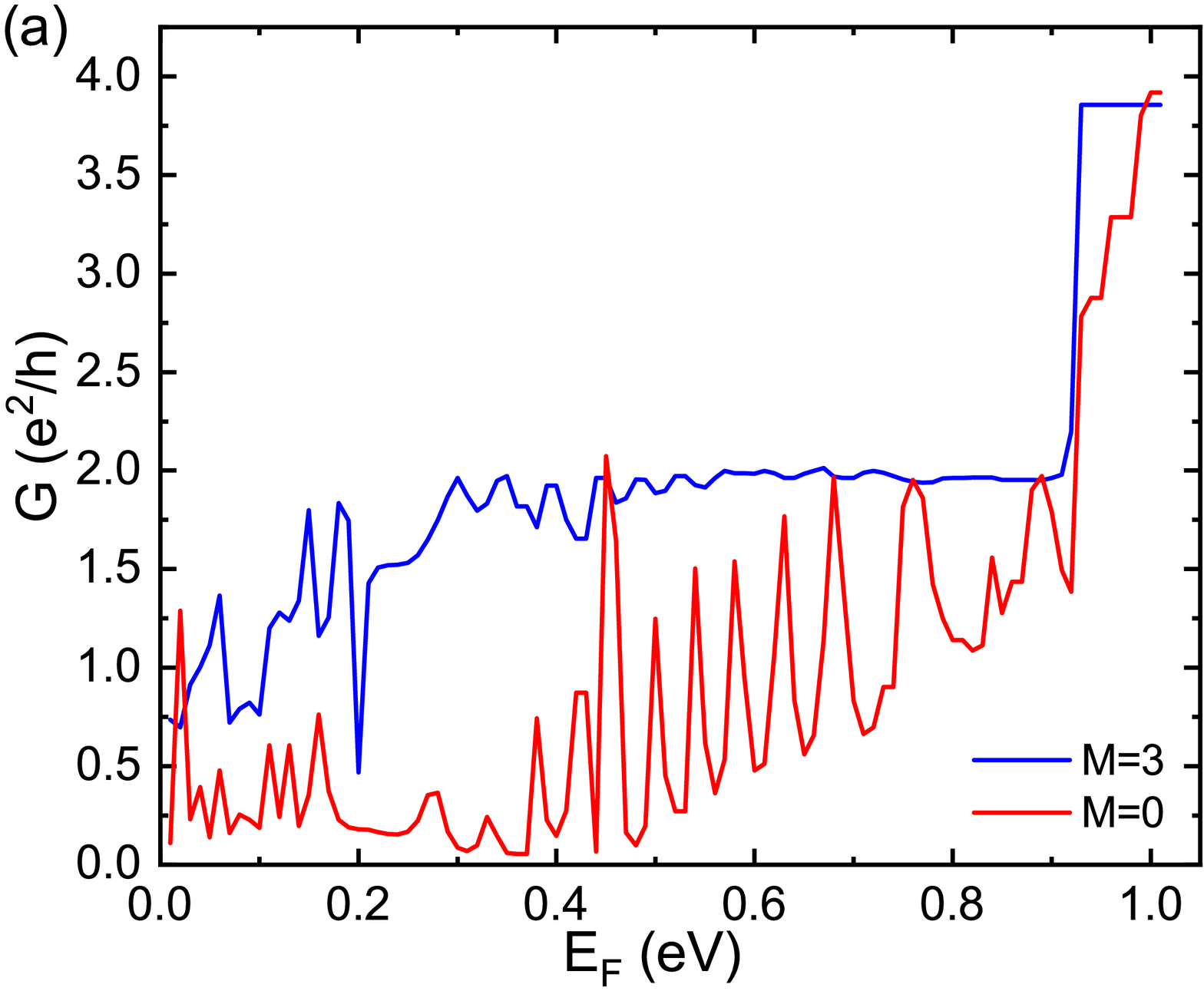}
\includegraphics[width=1\linewidth]{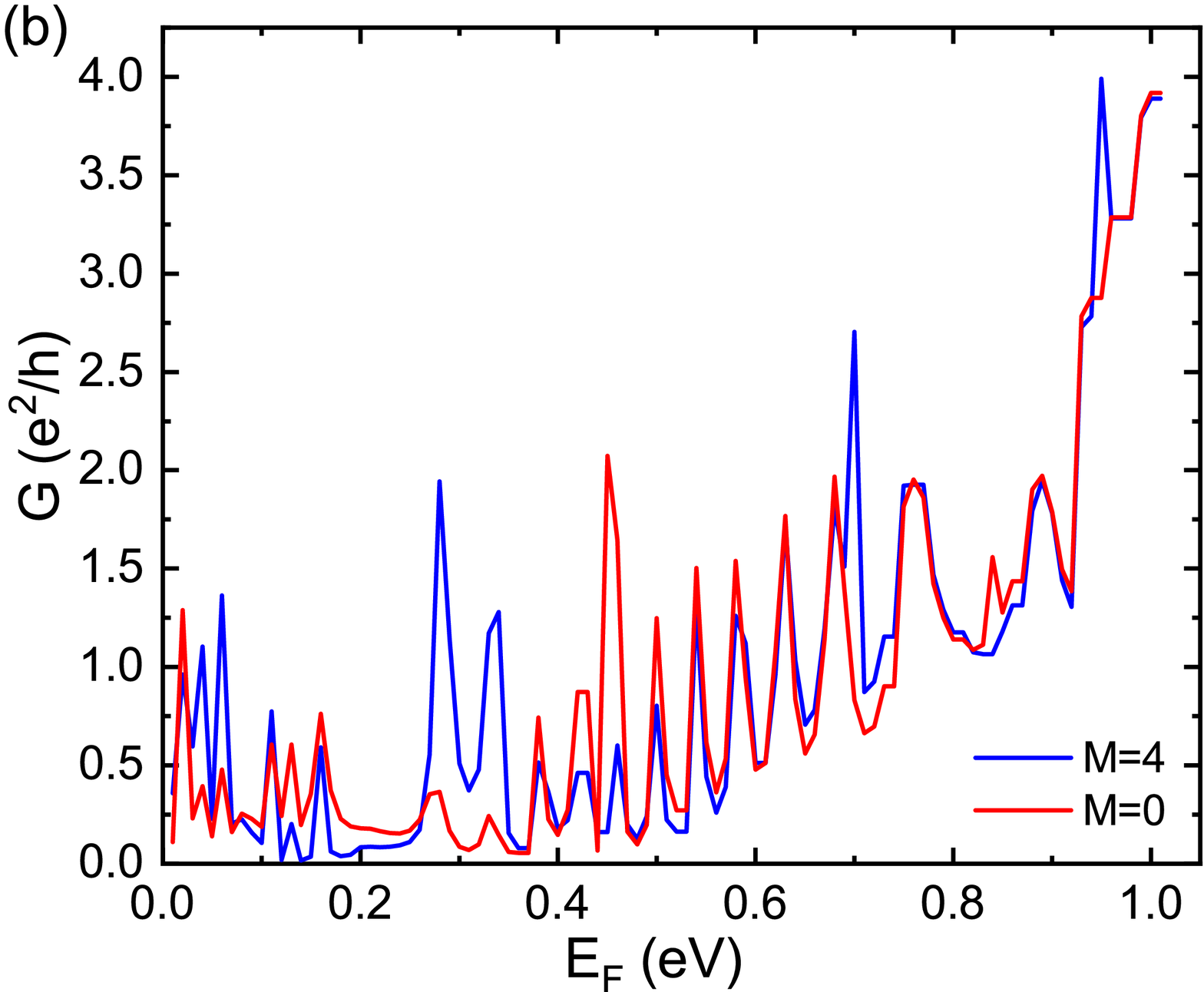}
\end{minipage}%
\begin{minipage}[b]{0.48\linewidth}
\includegraphics[width=1\linewidth]{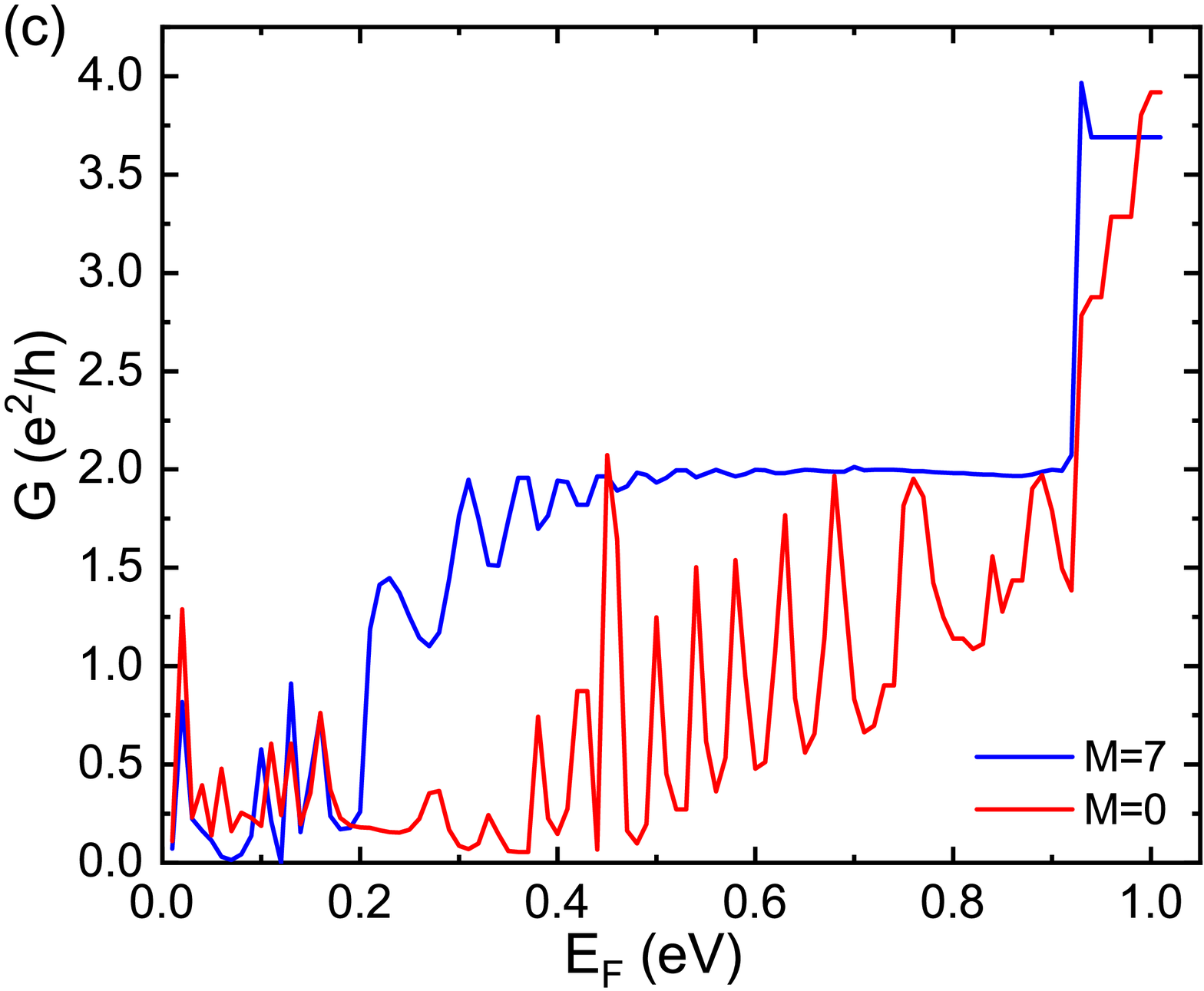}
\includegraphics[width=1\linewidth]{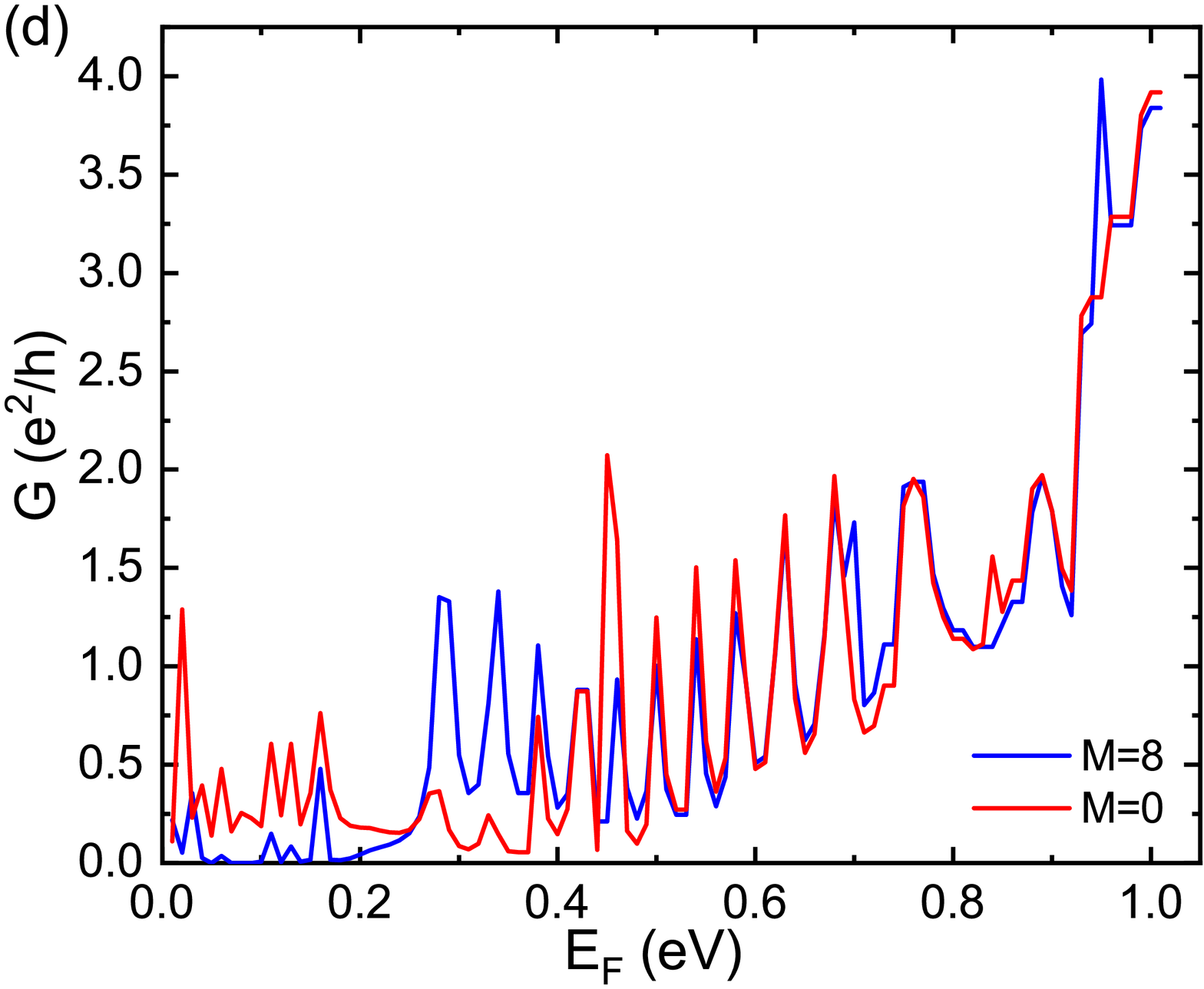}
\end{minipage}%
\caption{The conductance $G$ for different Fermi energy without external electric field and potential energy when the relative shift of the central part is (a) M=3 sites (b) M=4 sites (c) M=7 sites (d) M=8 sites. The sites number of the central nanoribbon is $N=4$.}
\label{fig:G0000}
\end{figure}

Next, we will study the influence of position of the central nanoribbon on the transport properties with and without external field effects. In Fig.~\ref{fig:G0000}, The conductance is plotted as a function of Fermi energy $E_F$ when the central narrow part is at different positions without the electric field and potential energy. As shown in Fig.~\ref{fig:move structure}, the relative distance of central scattering region for different geometrical structure is $\Delta y=M\xi$ with M being an integer. The width of central nanoribbon is 4 sites, that is $N=4$. When the relative shift is 3 sites (M=3), the conductance first vibrates and soon becomes steady after the Fermi energy reaches 0.5eV.
There is a step near $0.94~\mathrm{eV}$ and forms a new conductance plateau with $G\approx 3.75~\mathrm{e^2/h}$.
It can be seen that the change trends of the overall conductance is significantly different from the case where the nanoribbon does not shift (M=0).
When the relative shift is 4 sites (M=4), a stable plateau isn't produced.
There is an energy region with small conductance approaching to zero for the energy near $0.2$ eV, which can attribute to the interference of propagating modes
and the mode matching between the narrow and wide regions.

But while the central nanoribbon further shifts to 7 sites (M=7), the conductance has a stable value of 2$\mathrm{e^2/h}$ again between $0.5~\mathrm{eV}$ and 0.94eV of Fermi energy. However, this stable plateau disappears when we adjust the scattering area to be 8 sites (M=8) off the center.
The transport property basically keeps unchanged when the sites number $M$ is even. Similarly, the interference and the mode-matching effect induce zero conductance within the energy regime $0\sim 0.2$ eV, as shown in Fig.~\ref{fig:G0000}(d).
When the relative shift is even, there exist oscillating curves between the energy regime $0.5\sim 0.94$ eV, which arise from interference of the left-going modes and right-going modes. However, when the relative shift of the narrow part is odd (M=3 or 7), the interference is broken due to the asymmetrical structure and the oscillating peaks disappear. Accordingly, the conductance becomes a stable plateau between $0.5 \sim 0.94$ eV.
It is interesting that the constriction has different transport properties between odd- and even-site shift from the center. We also conducted a series of calculations which M has other values and attained the same phenomenon. These results show that the central nanoribbon has some different transport properties between even-M positions and odd-M positions. Some researches have been published that zigzag silicene nanoribbons with even widths and odd widths have very different current-voltage relationships, magnetoresistance effect and thermopower behavior, which can be attributed to the different parity of their $\pi$ and $\pi^\star$ bands under $c_2$ symmetry operation with respect to the center axis~\cite{kang2012symmetry,zhou2015symmetry}. However, our results further point out that the even-M positions and odd-M positions of central nanoribbons with the same width also have very different transport properties. Similarly, odd-M geometrical structure has lower symmetry, thus $\pi$ and $\pi^\star$ bands have no definite parity and contribute a larger conductance when electrons are injected from the wide part into the narrow constriction.
\begin{figure}[t]
\centering
\begin{minipage}[b]{0.48\linewidth}
\includegraphics[width=1\linewidth]{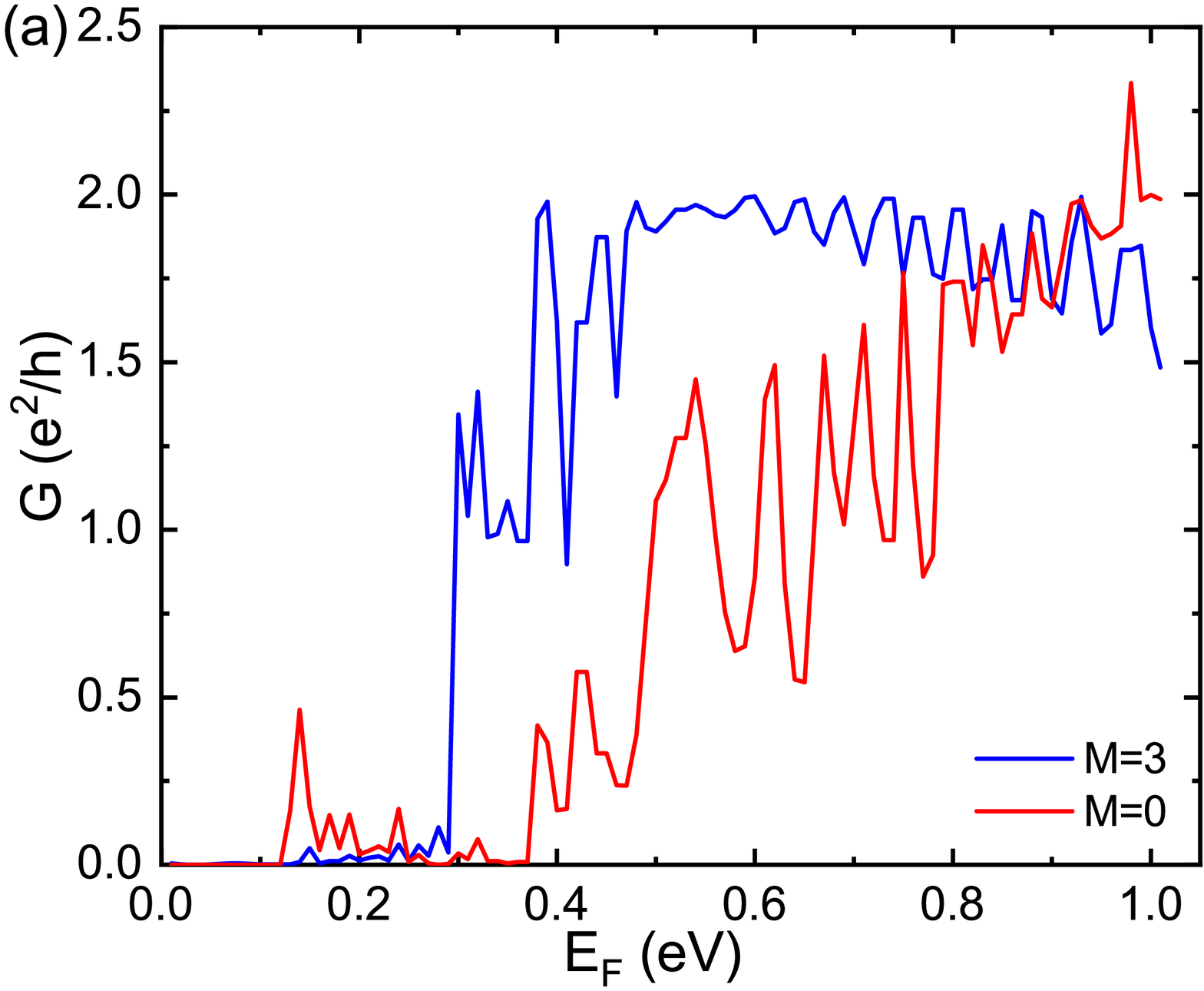}
\includegraphics[width=1\linewidth]{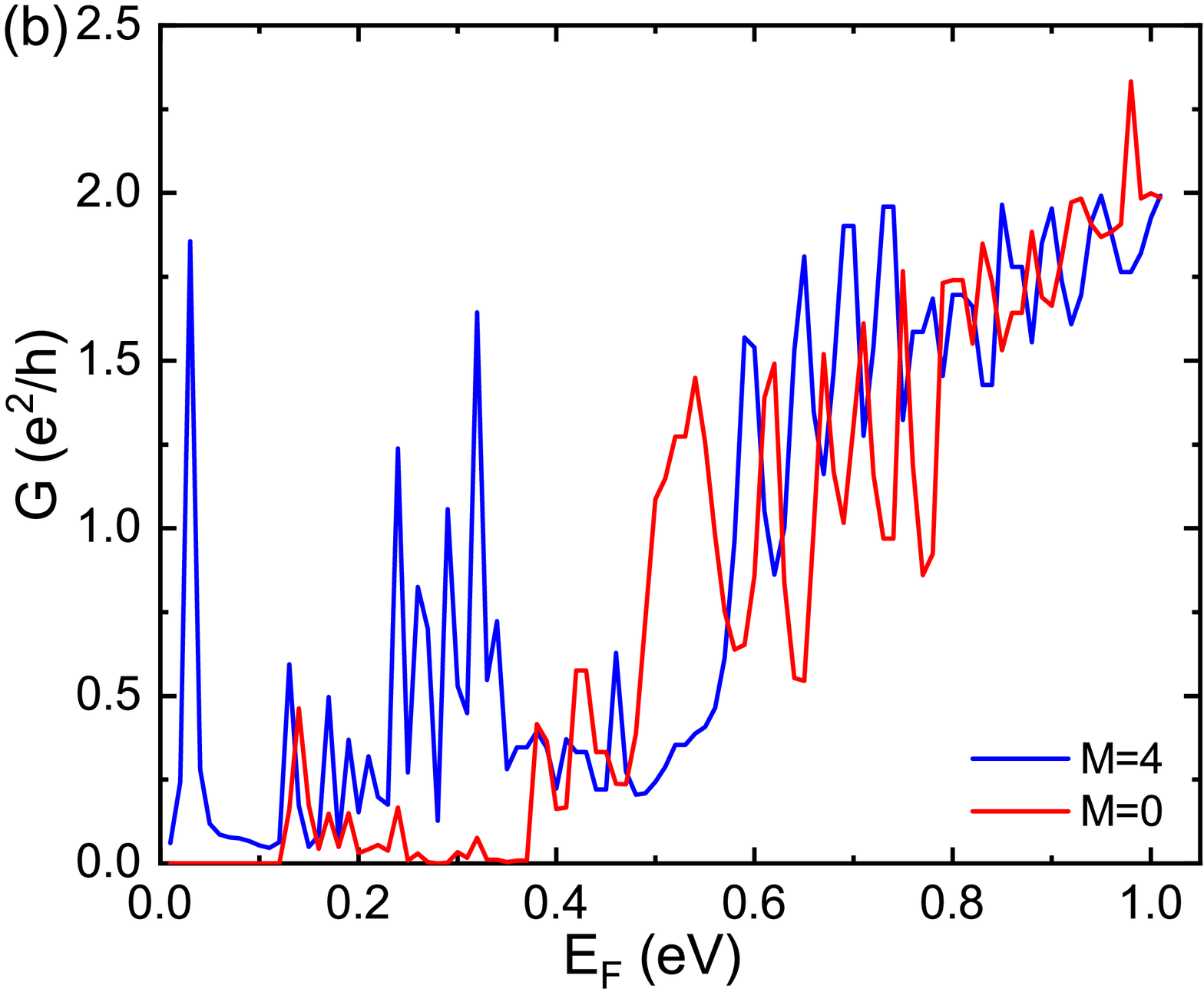}
\end{minipage}
\begin{minipage}[b]{0.48\linewidth}
\includegraphics[width=1\linewidth]{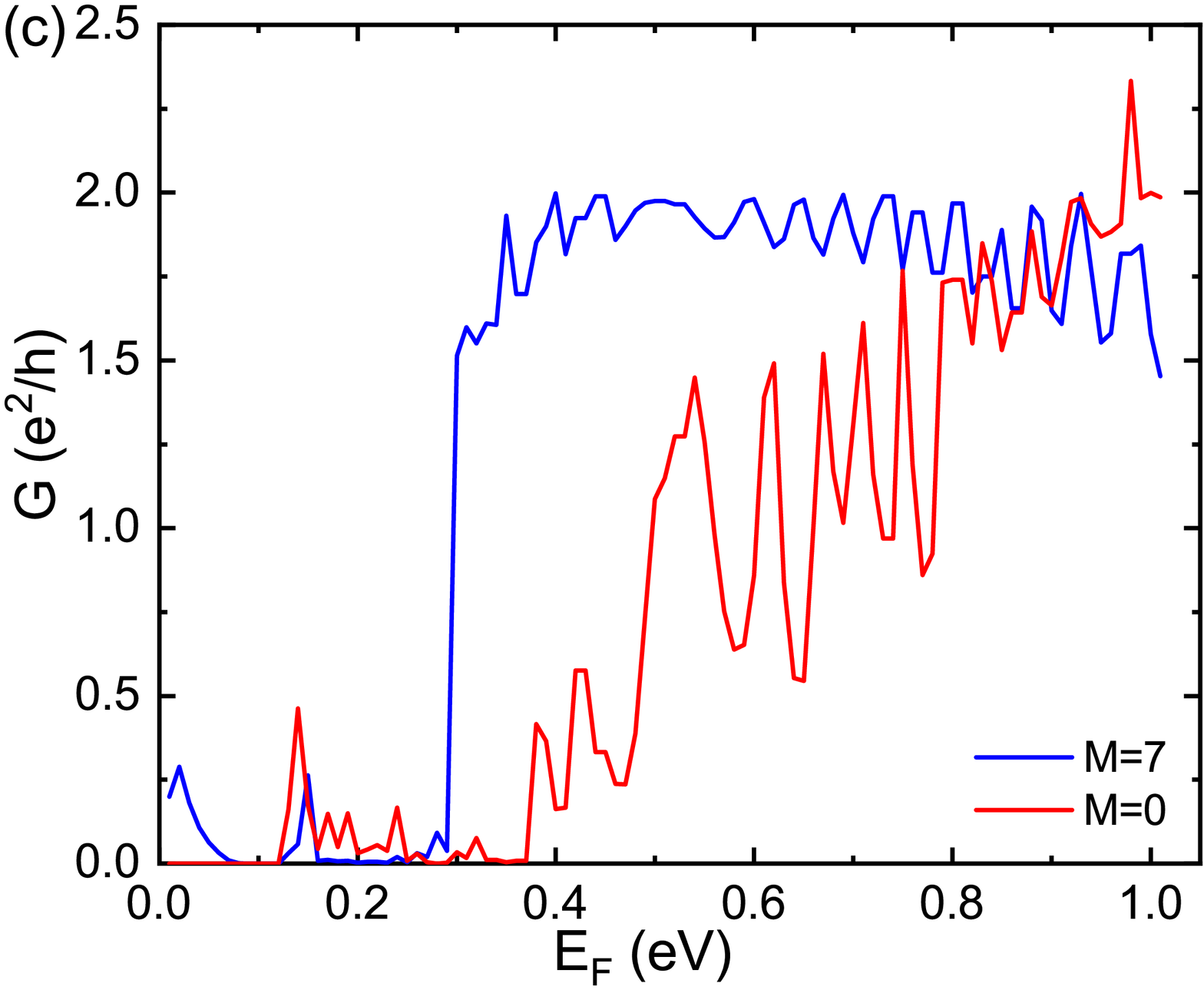}
\includegraphics[width=1\linewidth]{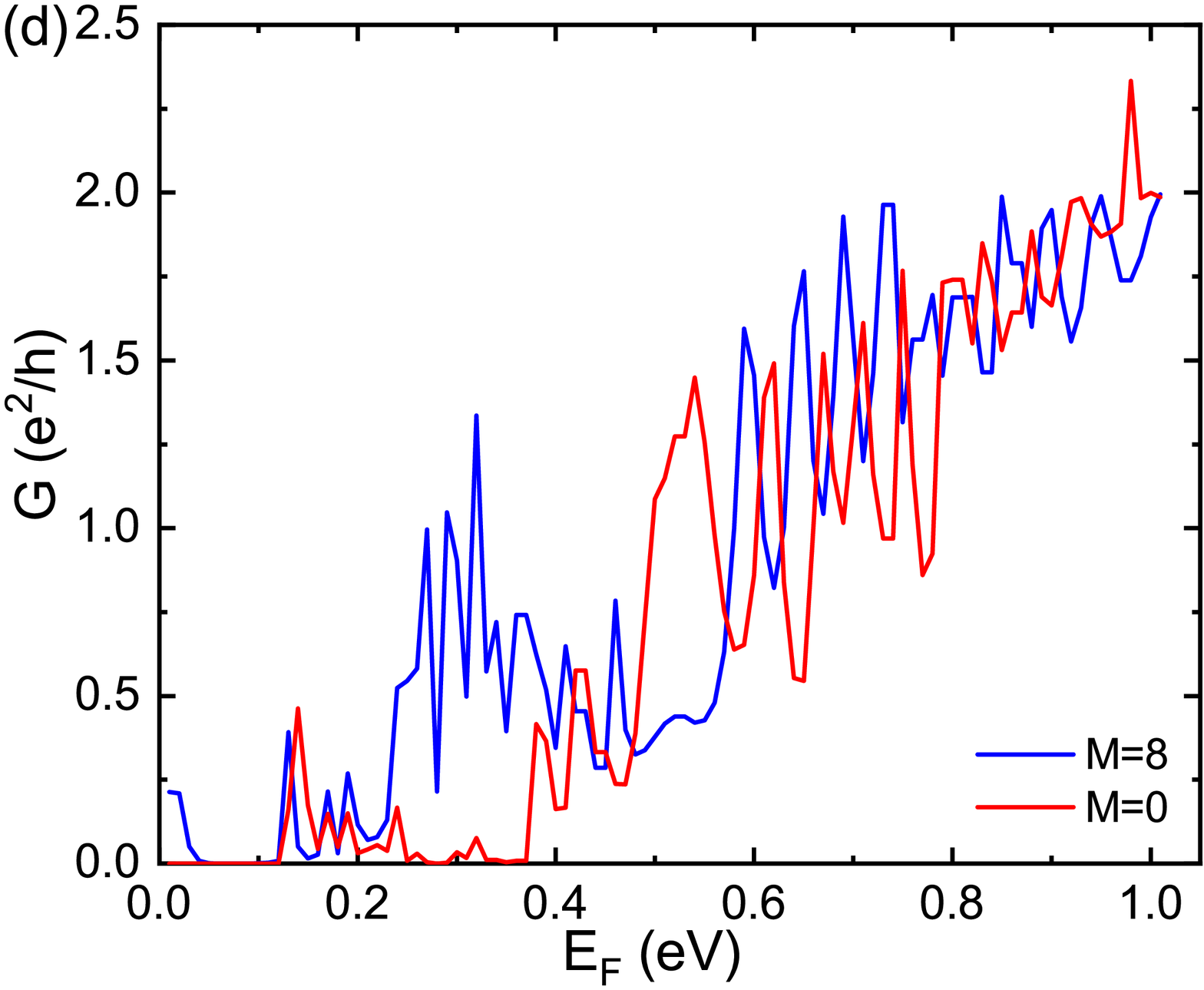}
\end{minipage}%

\caption{ The conductance $G$ for different Fermi energy with external electric field $E_z=0$ and potential energy $V_0=0.3~\mathrm{eV}$ when the relative shift of the central part is (a) M=3 sites (b) M=4 sites (C) M=7 sites (d) M=8 sites. The sites number of the central nanoribbon is $N=4$.}
\label{fig:G0300}
\end{figure}

Next, we investigate the effect of potential energy on the conductance of silicene device when its central nanoribbon also has a deviation from the central position. According to the results in Fig.~\ref{fig:G0300}, we could discover that the conductance have a notable difference with the patterns without potential energy. The conductance approaches to zero within the energy interval $0<E<0.12$ ($M=0$), which can be understood from the band structure as shown in Fig.~\ref{fig:band energy}. We assume that the electrons move from the left lead to the central scattering region. In the energy interval $0<E<0.12$, there exists only one conduction subband supporting the right-moving electrons of valley K' (with longitudinal wavevector $k_xa\in(\pi,2\pi)$) in the leads. While in the central nanoribbon, the highest valence subband only permits the right-moving holes in the same energy valley K', which results in a zero conductance~\cite{li2019electrically}. With increasing Fermi energy, more subbands in leads permit left-moving electrons to flow into the central scattering region to match the right-moving propagating modes of holes, which induces small value of the conductance [see red curve of Fig.~\ref{fig:G0300}].
When M is odd, the conductance is almost zero until the Fermi energy up to $0.3~\mathrm{eV}$. Additionally, there is a sudden promotion around $0.3~\mathrm{eV}$ of Fermi energy, which makes the pattern like a step. It is interesting as it may be used as a switch. But if M is even, the step disappears. The conductance is relatively unstable and tends to raise with increasing of Fermi energy.
\begin{figure}[t]
\centering
\begin{minipage}[b]{0.48\linewidth}
\includegraphics[width=1\linewidth]{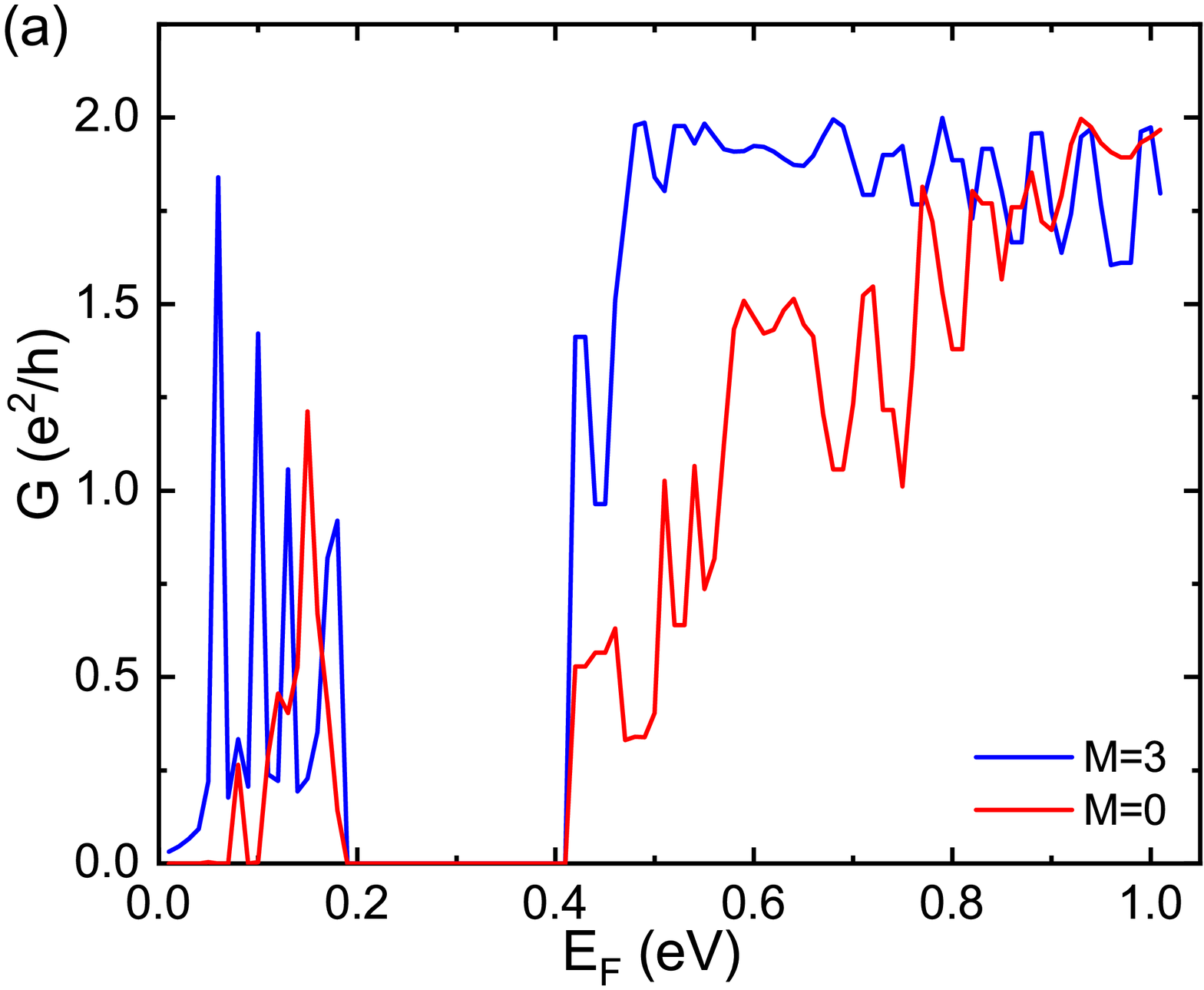}
\includegraphics[width=1\linewidth]{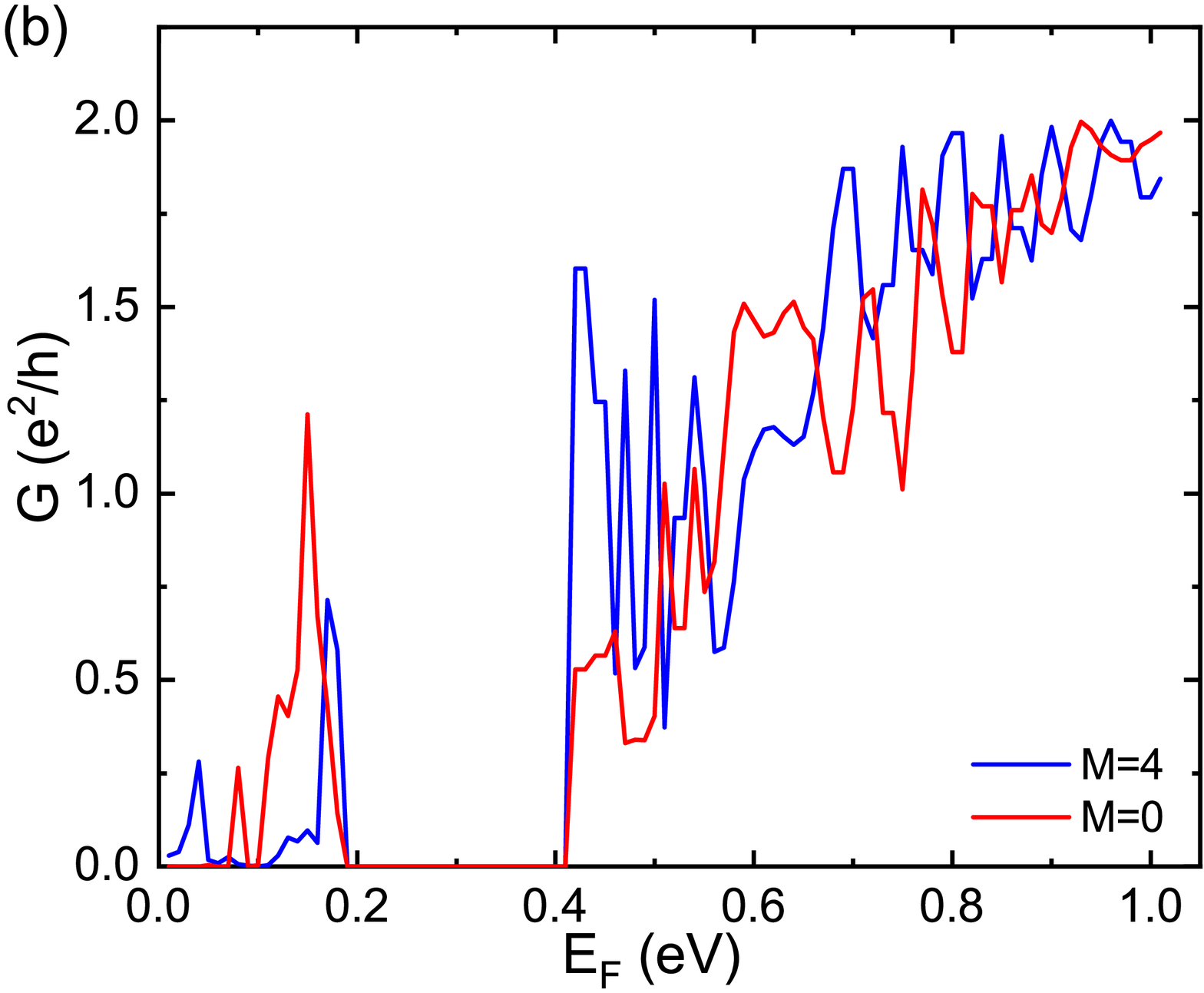}
\end{minipage}%
\begin{minipage}[b]{0.48\linewidth}
\includegraphics[width=1\linewidth]{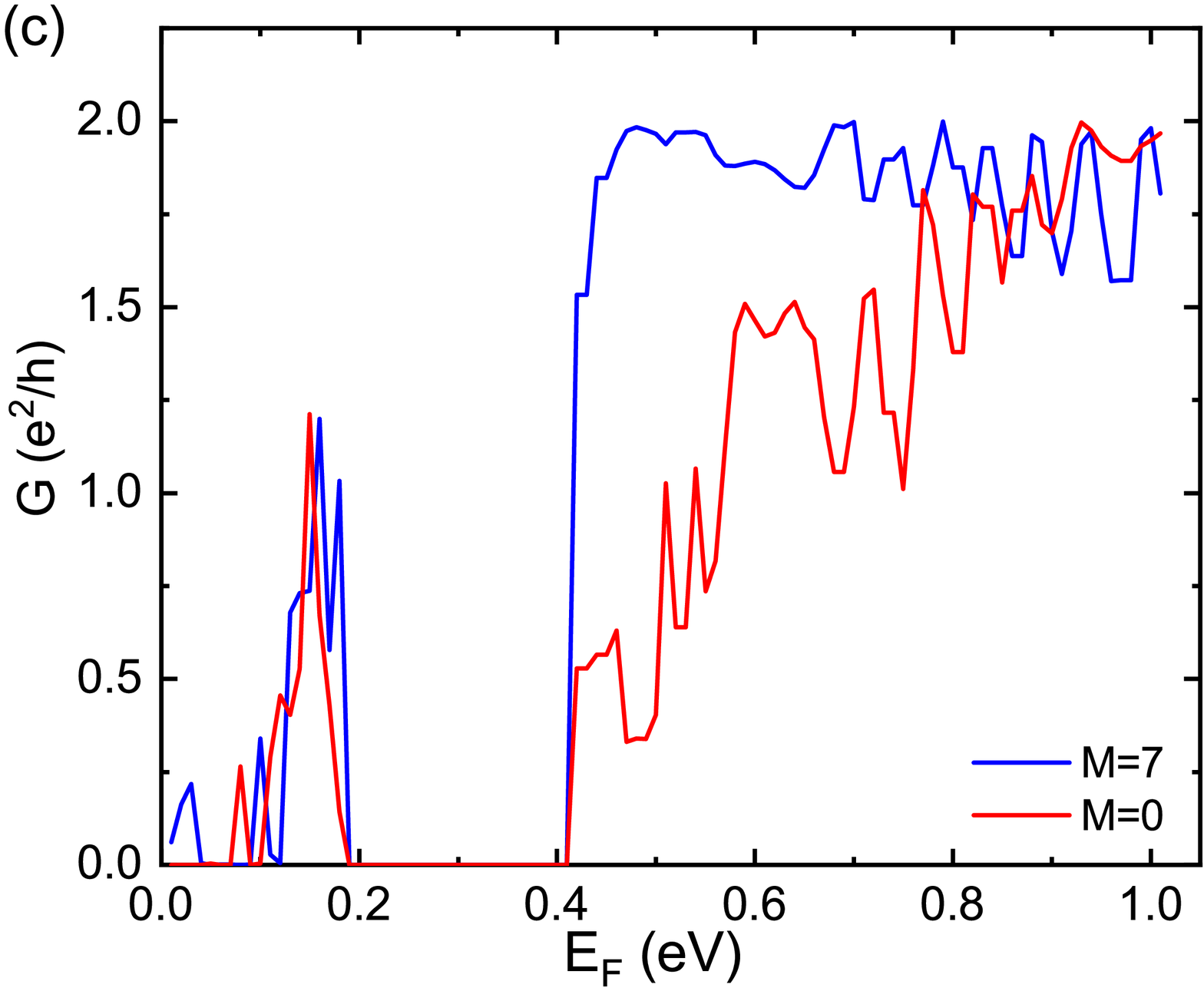}
\includegraphics[width=1\linewidth]{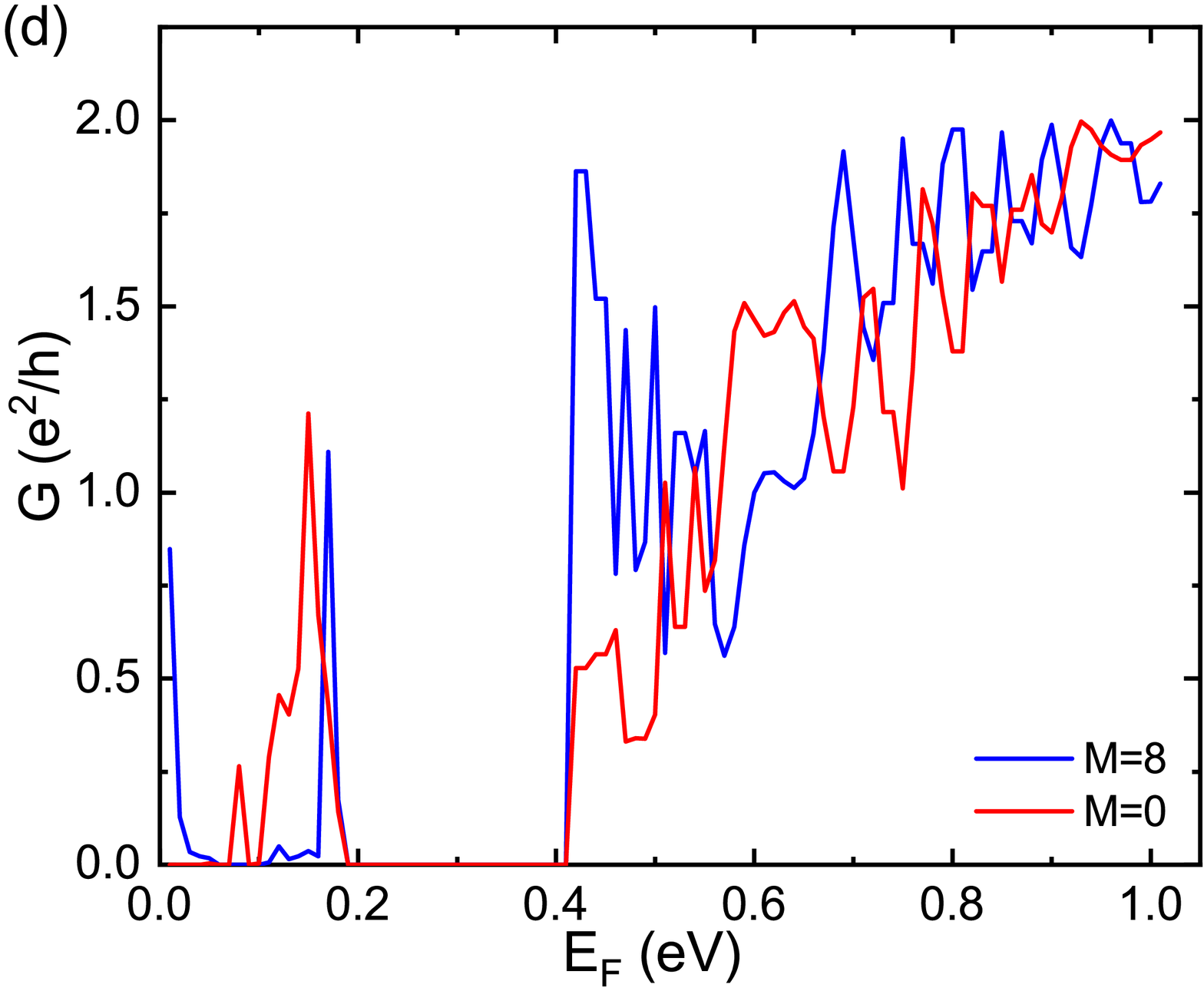}
\end{minipage}%
\caption{ The conductance for different energy with external electric field $E_z=0.5~\mathrm{eV{\AA}^{-1}}$ and potential energy $V_0=0.3~\mathrm{eV}$ when the relative shift of the central part is (a) M=3 sites (b) M=4 sites (C) M=7 sites (d) M=8 sites. The sites number of the central nanoribbon is $N=4$.}
\label{fig:G0305}
\end{figure}

Furthermore, we investigate the combined influence of external electric field and the potential energy on silicene constrictions. By adding an electric field of $0.5~\mathrm{eV{\AA}^{-1}}$ to the device, we use the same way to study the transmission property. As shown in Fig.~\ref{fig:G0305}, the conductance has several peaks when the Fermi energy is below $0.19~\mathrm{eV}$. And then, the conductance remains zero between $0.19~\mathrm{eV}$ and $0.41~\mathrm{eV}$ of Fermi energy in all the patterns of Fig.~\ref{fig:G0305}, which forms a sharp contrast with the patterns in Fig.~\ref{fig:G0000} and Fig.~\ref{fig:G0300}. Apparently, this is because the external electricity opens the energy band gap, as shown in Fig.~\ref{fig:band energy}(b). When the Fermi energy is larger than $0.41~\mathrm{eV}$, these figures show the difference between even-M and odd-M cases again. The conductance has an apparent step when the relative shift is odd sites, while the patterns of even-M shift have a decrease of about 50\% and greater fluctuations before gradually raising and stable around 1.75 $\mathrm{e^2/h}$ forming a pothole-like pattern. Although the conductances of the four patterns all oscillate, the oscillating amplitude is significantly larger when the value of M is even. What's more, the overall conductance of the constriction with odd value of M is bigger than that of the even-site shift case.

\begin{figure}[t]
\centering
\begin{minipage}[b]{0.48\linewidth}
\includegraphics[width=1\linewidth]{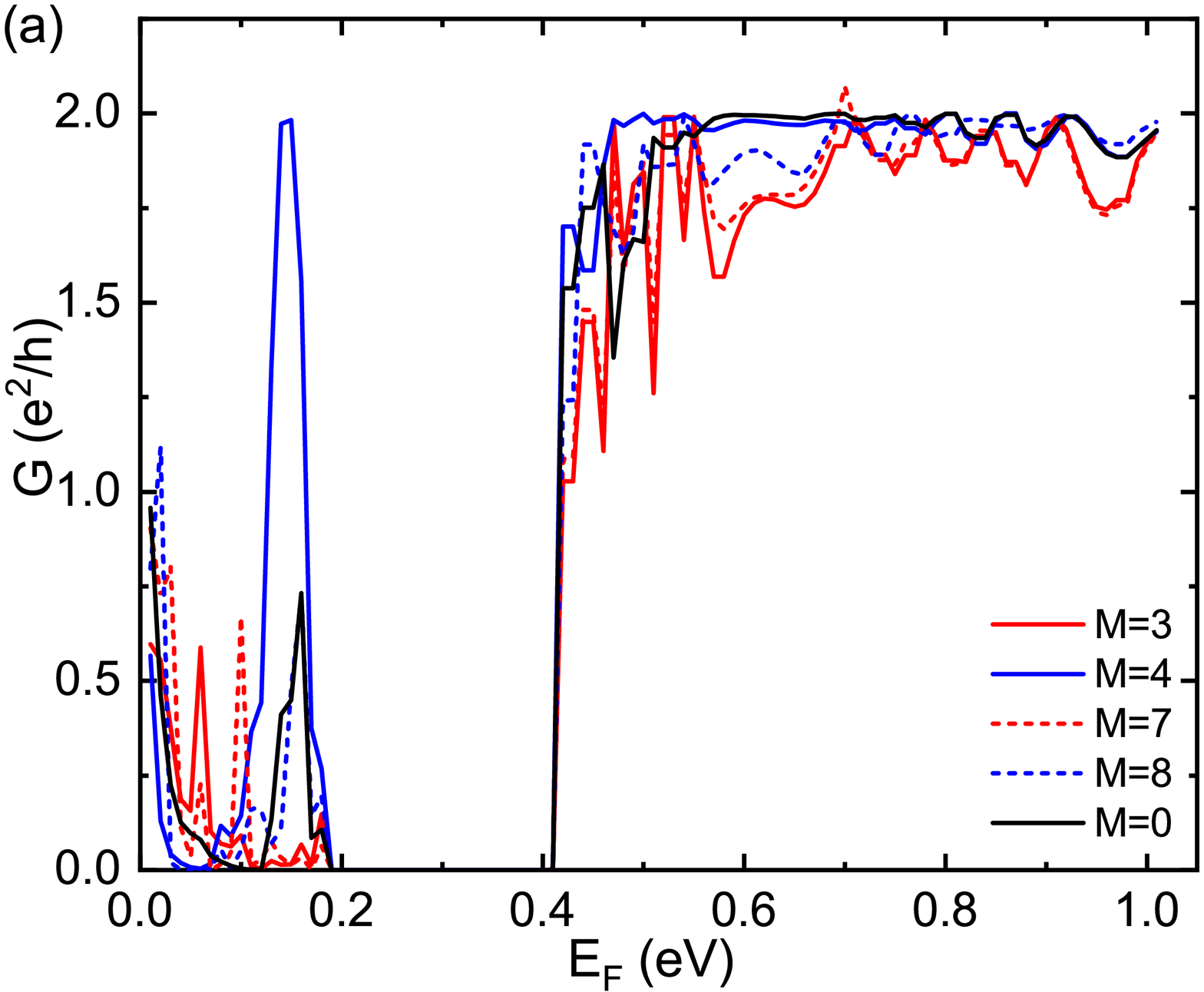}
\includegraphics[width=1\linewidth]{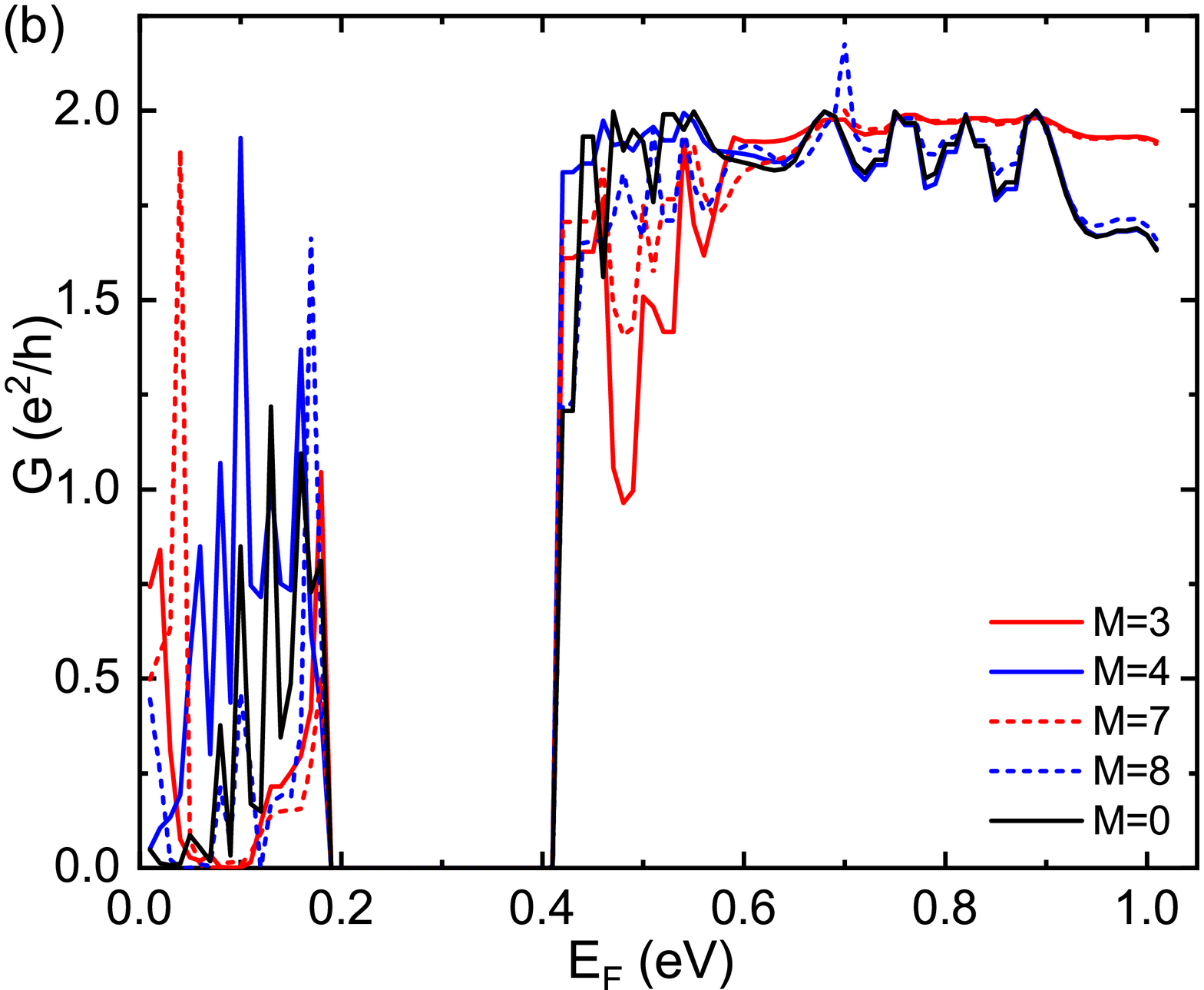}
\end{minipage}%
\begin{minipage}[b]{0.48\linewidth}
\includegraphics[width=1\linewidth]{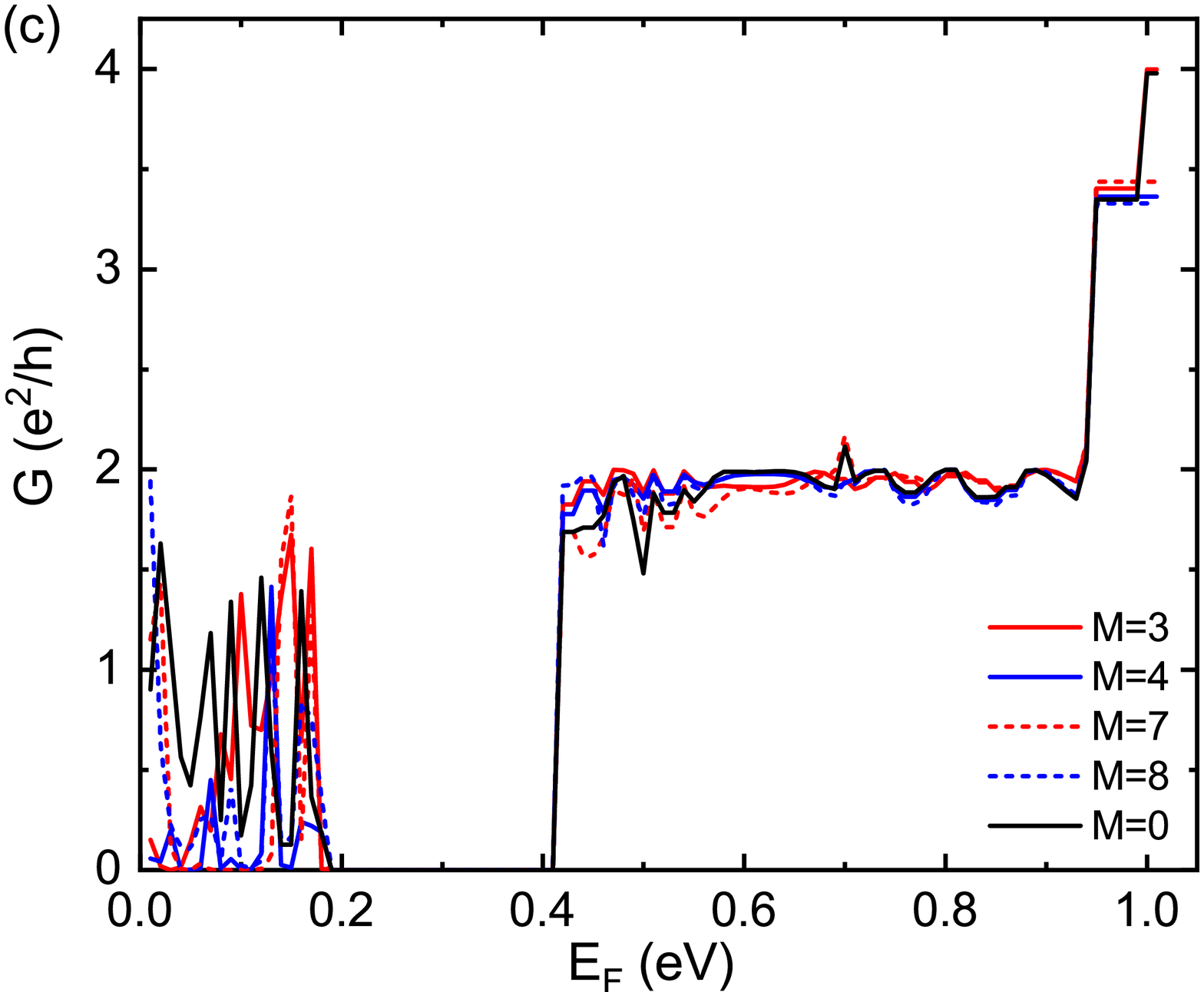}
\includegraphics[width=1\linewidth]{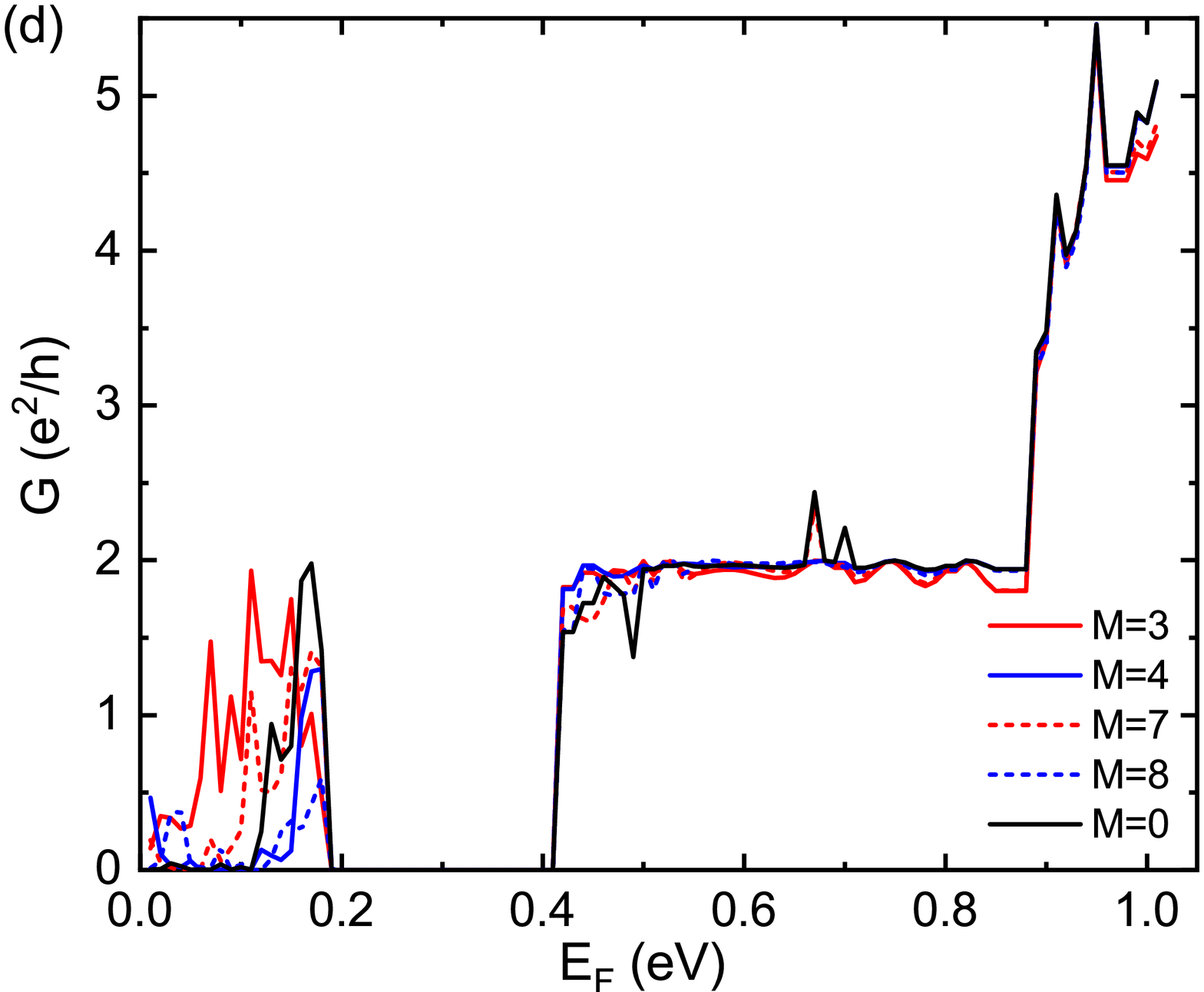}
\end{minipage}%
\caption{ The conductances are plotted as a function of Fermi energy by using different widths and shifting distances of the central nanoribbon. The width is (a) 5 sites (N=5), (b) 6 sites (N=6),  (C) 7 sites (N=7) and (d) 8 sites (N=8). The other parameters are $E_z=0.5~eV{\AA}^{-1}$ and $V_0=0.3eV$.}
\label{fig:different width}
\end{figure}
In order to further study the influence of central scattering region on the transport, we change the width of central nanoribbon and then test its transmission with the electric field $E_z=0.5~\mathrm{eV{\AA}^{-1}}$ and potential energy $V_0=0.3~\mathrm{eV}$. As we broaden the width of the scattering region, we found that the characteristic of the transport has some important changes. The most intuitive change is that all the conductance patterns have the "step" shape regardless of whether M is odd or even. We noted that the deviation of the odd-even sites seems to have a reversed effect on the conductance. As shown in Fig.~\ref{fig:different width}, the steps are more stable in even-M cases instead. And it can be seen that when the width is 5 and 6 sites (Fig.~\ref{fig:different width}(a), (b)) with odd-M, the figures show the pothole-like patterns which appear in Fig.~\ref{fig:G0305} where the width is 4 sites and M is even instead. And as the width increases, the pothole-like pattern becomes narrower and shallower, and finally a step pattern similar to the odd-M case can form. This illustrates that odd-even effect of M is weakened as the width increases. In addition, when Fermi energy is greater than $0.41~\mathrm{eV}$, the conductance of the silicene heterojunction becomes more stable as the width increases.

But how to explain the phenomenon that the conductance of the system with odd-M and even-M shift reverses? After analysis, we found that when the width of the central nanoribbon is 4 sites, the distance between the top edges of the wide and narrow parts (marked as $\Delta h$) has 18 sites when $M=0$. So when M is odd, $\Delta h$ has odd sites. For example, when $M=3$, $\Delta h=15$ sites. However, when the site number is $N=5$, the narrow region has 3 sites above the transverse center and 2 sites beneath the center. So when the shift parameter M is odd, $\Delta h$ has even sites. When $N=6$, the narrow region has 3 sites above the transverse center and 3 sites beneath the center. Similarly, when M is odd, $\Delta h$ has also even sites. Since we only consider the upward shift of the central region, its width above the transverse center combined with the shifting sites will decide the parity of $\Delta h$. Seen from above results, we found that the parity of $\Delta h$ is consistent with the changing curves of the conductance. It is reasonable to speculate that the different behaviour of the conductance is actually related to the odd and even-site numbers of $\Delta h$. In order to verify our speculation, we adopt the structure with 2 sites of width above the transverse center and 3 sites beneath the center to calculate the transport property. The obtained results show that when the site number of $\Delta h$ is odd, the platform of the step-like pattern of the conductance is indeed more stable. As the width of the central region increases, the difference of the parity gradually decreases, and the patterns have steps. Therefore, it can be further revealed from the figure that when the width D is small, the conductance of odd-$\Delta h$ structure is more stable than the even-$\Delta h$ case. As the width D of the central region increases, the pothole pattern gradually disappears and the odd-even effect of $\Delta h$ disappears. For convenience of expressing the position-dependent relation, we still use the parameter $M$ to describe our results in the following passages.
\begin{figure}[t]
\centering
\includegraphics[scale=0.58]{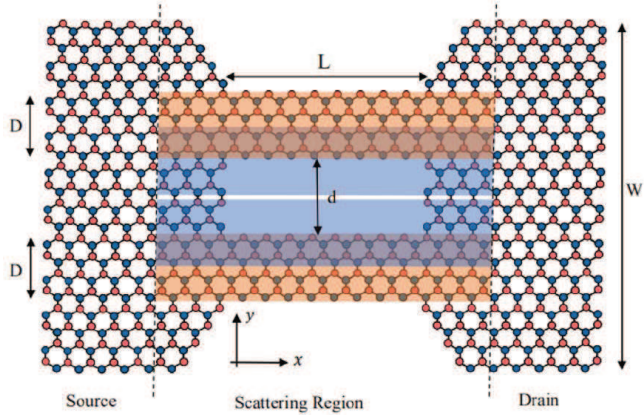}
\caption{Schematic diagram of the silicene constriction with two narrow ribbons in central scattering region. The zigzag direction is parallel to $x$ axis as shown in the picture. The variety of blue and orange region shows two different distances of the two narrow ribbons. $d=2M\xi$ is defined as the distance between two ribbons.}
\label{fig:two tunnels}
\end{figure}

\begin{figure}[t]
\centering
\includegraphics[scale=0.35]{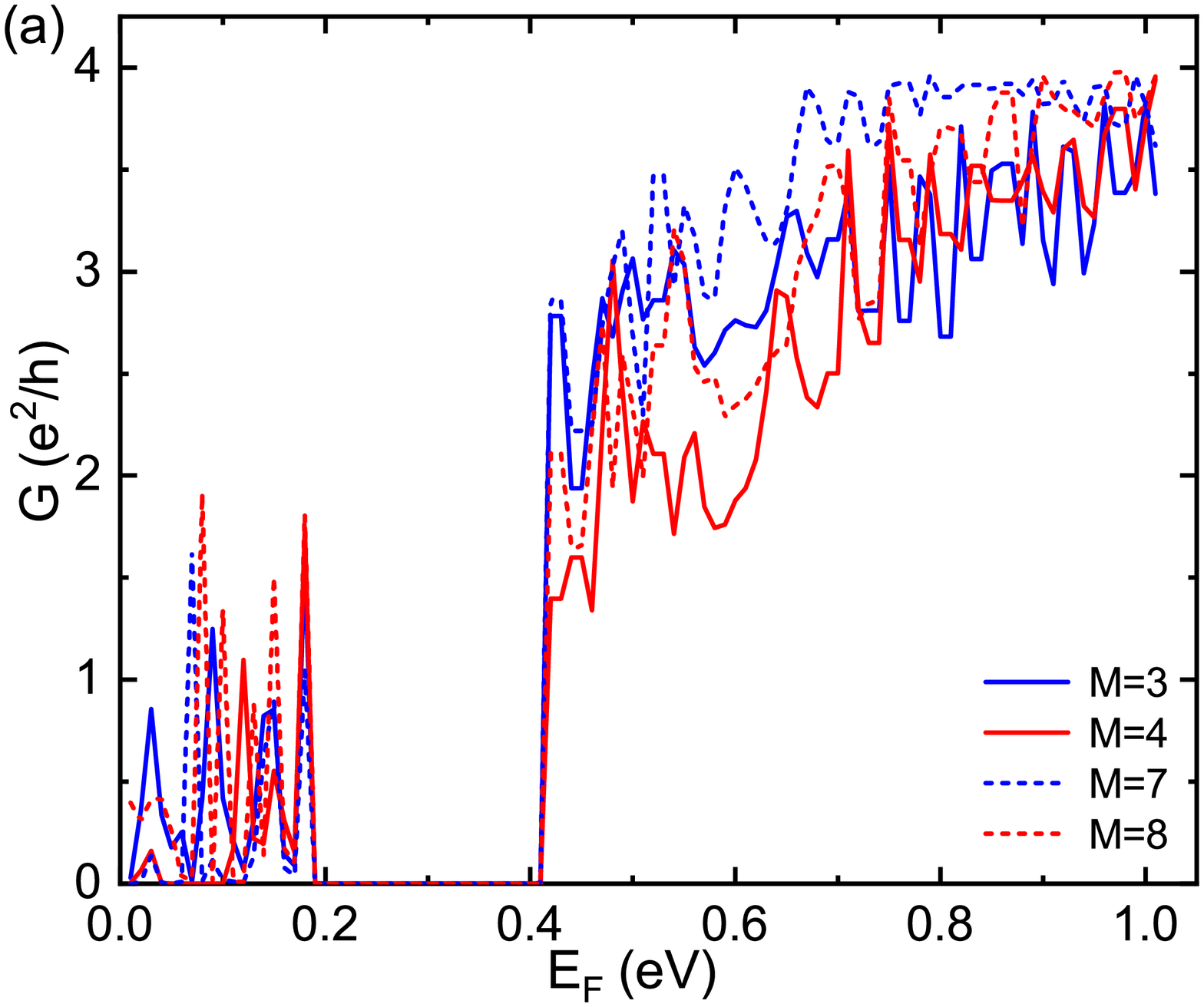}
\includegraphics[scale=0.35]{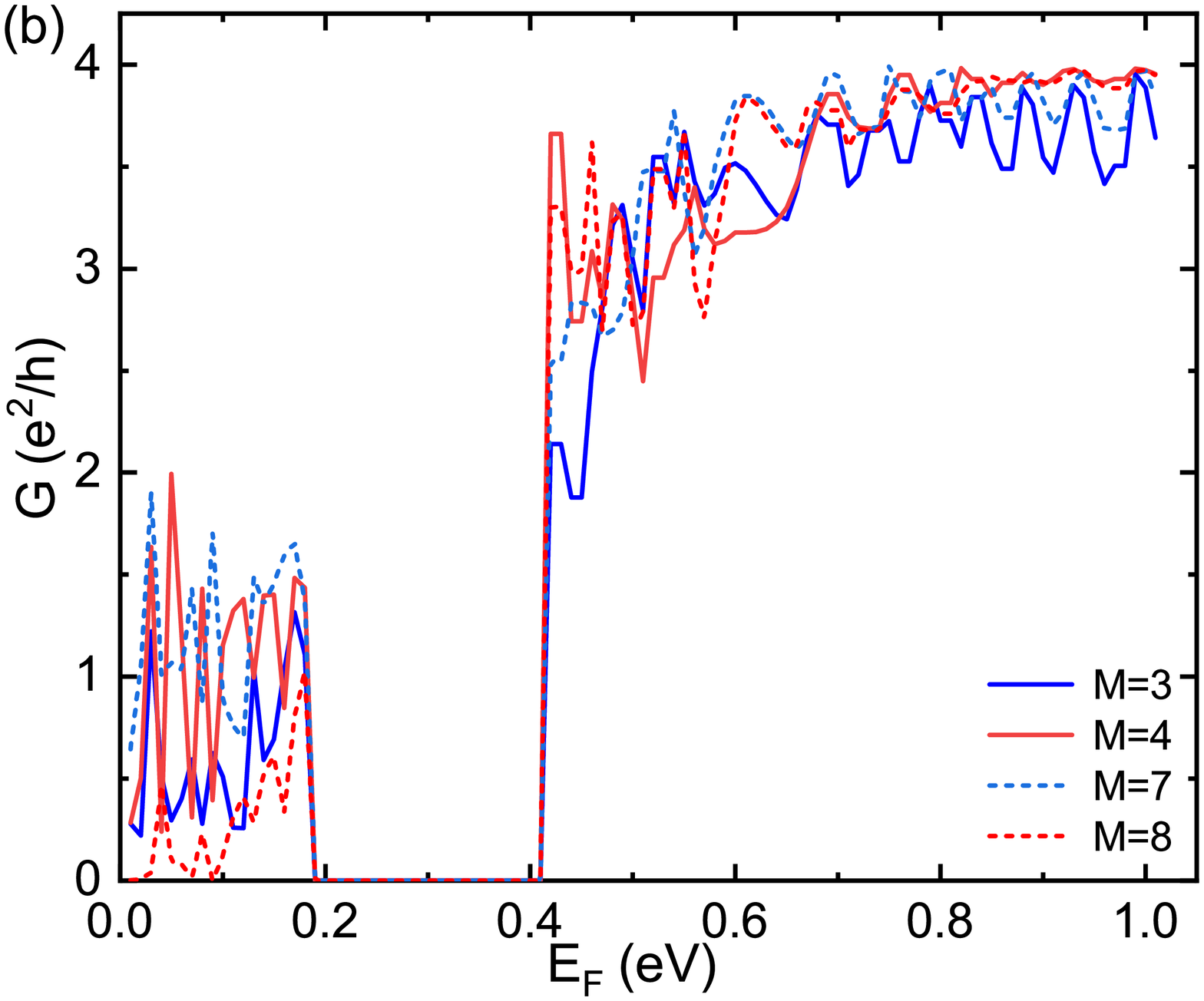}
\caption{ The conductances are plotted as a function of Fermi energy in the structure of two channels, and the width of each channel is (a) 3 sites ($N=2\times3$), (b) 4 sites ($N=2\times4$). The other parameters are $E_z=0.5~\mathrm{eV\AA}$ and $V_0=0.3eV$.}
\label{fig:two bands}
\end{figure}

At last, we develop a novel structure of the central scattering region and investigate its transmission. As shown in Fig.~\ref{fig:two tunnels}, the central region are divided into two symmetrical parts along the transverse direction. When the two nanoribbons are separated, the total width of the central region is $2D=(2N-2)\xi+\frac{1}{2}a$, in which $2N=4, 6, 8,...$ is the total sites number of the two nanoribbons. We also consider the situation of different distances between the two channels. The distance between two channels is $d=2M\xi$. When $d=0$, as the blue region shows, the case is the same as in the one-channel constrictions above whose width of central nanoribbon is $2D$, in which $2D=(2N-1)\xi+\frac{1}{4}a$. Here, we only consider the case of two nanoribbons with structures that are symmetrically shifted from the original position of $d=0$ to the positive and negative y-axis. By expanding the one channel of scattering region to two channels, we get a totally different transport property. The results are shown in Fig.~\ref{fig:two bands}.

It can be seen that in this structure, the entire silicene heterojunction is symmetrical in the transverse direction. We choose the case that the total width is $N=2\times3$ and $N=2\times4$ for analysing. The width of each channel of the nanoribbon is 3 sites in Fig.~\ref{fig:two bands}(a) and 4 sites in Fig.~\ref{fig:two bands}(b). Under the conditions of potential energy $V_0=0.3~\mathrm{eV}$ and the electric field $E_z=0.5~\mathrm{eV{\AA}^{-1}}$ for these two-channel silicene constrictions, the distance between the two channels is changed for numerical calculation. Assuming that the initial distance between the two channels is zero, $M$ represents that the two channels are simultaneously shifted by M sites along the positive and negative y-axis. For example, when $M=3$, the two channels move up and down 3 sites respectively, so the distance between them is 6 sites. As shown in Fig.~\ref{fig:two bands}, the two-channel constrictions have similar "step" shapes with those of one-channel constrictions. But the maximum value of the conductance in two-channel silicene is almost twice of that of one-channel case though their total width is equal. For example, the maximum conductance of the center scattering region with $N=8$ (shown in Fig.~\ref{fig:different width}(d)) is only about half of the value in the two-channel structure with $N=2\times4$ (shown in Fig.~\ref{fig:two bands}(b)). Apparently, this two-channel structure has an excellent performance on the conductance.
From the experimental point of view, the narrow silicene constriction with a relative shift of the central region can be fabricated by covering some areas to avoid deposition of silicon atoms. The two-channel structure can also be fabricated by using the same method in experiments.

\section{Conclusions}
In a conclusion, we mainly studied the influence of the geometric structure of the silicene constriction on the transport properties. By calculating the relationship between the conductance and Fermi energy, we have systematically studied how the potential energy, the applied electric field, and the geometry of the central scattering region regulate the transport properties. The results show that when there is no potential energy and electric field, the odd-sites shift of the central nanoribbon will make the conductance have a stable area, while the even-sites shift will not produce such a "platform" area. When the effect of potential energy is added, the conductance of the Fermi energy less than 0.3eV is almost zero, and the "platform" area generated by the odd-sites shift makes the conductance produce an approximate "step" pattern, while the even-sites shift will not produce "steps". Next, we added an external electric field under the previous conditions, and found that the conductance is zero when Fermi energy is at the regime of $0.19-0.41~\mathrm{eV}$. The abrupt change of silicene conductance in the case of odd-$M$ produces a "step", while the silicene conductance in the case of even-$M$ oscillates and rises gradually. Then, we studied the conductance by the width of the scattering region. We found that the "step" is actually affected by the odd and even value of $\Delta h$. But the wider is the width, the smaller is the parity effect of $\Delta h$: the conductance becomes more stable in the case of odd-$\Delta h$, while the conductance in the case of even-$\Delta h$ gradually shows a step. The wider is the width of the nanoribbon, the more stable is the "step". However, it can be found from the results that although the conductance has some interesting changes under external regulation, the conductance "step" only rises to close to 2 $e^2/h$. At last, we promoted a novel two-channel structure and found that its conductance will be almost twice of the conductance in the one-channel structure, though they have the same total width.

\begin{acknowledgements}
This work was supported by National Natural Science Foundation of China (Grant No. 11574067).
\end{acknowledgements}

\end{document}